\documentclass[prb,twocolumn,floatfix,showpacs]{revtex4}

\usepackage[dvips]{graphicx}
\usepackage[dvips]{color}
\usepackage{hyperref}

\usepackage{amsmath}
\usepackage{amsfonts}

\newcommand{\crt}[1]{\hat{c}^{\dagger}_{#1}}
\newcommand{\ann}[1]{\hat{c}^{ }_{#1}}
\newcommand{\hcrt}[1]{\hat{b}^{\dagger}_{#1}}
\newcommand{\hann}[1]{\hat{b}^{ }_{#1}}
\newcommand{\bra}[1]{\langle #1 |}
\newcommand{\ket}[1]{| #1 \rangle}

\newcommand{\eg}{e.g.\ }

\newcommand{\ie}{i.e.\ }
\newcommand{\etal}{et al.\ }

\newcommand{\mmax}[2]{\ifdim#1>#2#1\else#2\fi}
\newcommand{\mmin}[2]{\ifdim#1<#2#1\else#2\fi}

\begin{document}

\title{Disorder and Interactions in 1D Systems}
\author{Jonathan M Carter}
\author{Angus MacKinnon}
\affiliation{Blackett Laboratory, Imperial College London, South Kensington Campus, London SW7 2AZ, UK}
\email{a.mackinnon@imperial.ac.uk}  
\date{\today}

\message{\the\linewidth}
\begin{abstract}
We present a new numerical approach to the study of disorder and interactions
in quasi-1D systems which combines aspects of the transfer matrix method and
the density matrix renormalization group which have been successfully applied
to disorder and interacting problems respectively.  The method is applied to
spinless fermions in 1D and the existence of a conducting state is
demonstrated in the presence of attractive interactions.
\end{abstract}
\maketitle
\section{Introduction}
It is well established that in the presence of disorder electron wavefunctions
can become localized.  Considerable numerical work has been carried out for
non-interacting systems with results reaching a reasonable consensus: theory
and experiment are in general qualitative agreement.  However, in 3D the
calculated value of the universal critical exponent is markedly larger than
the empirically measured value \cite{slevin}.  This seems to suggest that an
essential factor is missing from calculations: the obvious candidate is the
electron-electron interaction.  Furthermore, some have claimed to observe a
Metal-Insulator transition in 2D contrary to the widely accepted scaling
theory of Anderson localization \cite{krav}.  This is often accredited to the effect of interactions.  Hence during the last 10 years attention has been switching to this more difficult case.  The central problem is that the model becomes a many-body system and so the Hilbert space grows quickly with system size.  This renders an exact numerical calculation far beyond computational capabilities.  Nevertheless, several studies have been accomplished, these suggest inclusion of interactions may yield non-trivial behavior.
 
Shepelyansky \cite{shep} performed calculations on two interacting particles.
In 1D, interactions caused a large enhancement of localization length.  Other
work showed that in 2D the effect is possibly stronger leading to
delocalization \cite{ortuno}.  However, some caution is required as the method fails to reproduce known non-interacting results when interactions are switched off.
 
The most successful method for treating the finite density problem is the
Density Matrix Renormalization Group (DMRG) approach \cite{white2,white5}. 
This works by performing a direct diagonalization but reducing the Hilbert space by systematically
discarding basis states that do not contribute significantly to the ground
state.  Applying this method to the Anderson interacting model (defined in
equation \ref{eq:ham}), a delocalized regime was found for attractive
interactions \cite{schmit4}.  In more recent papers by the same authors, it was
noted that interesting physics is washed out in the averaging process.  Charge
reorganizations can be seen as electrons on a chain shift from the Mott
insulator limit (strong interactions) to the Anderson insulator limit
(strong disorder) \cite{schmit6}.  Extensions of DMRG to 2D have encountered difficulties.
 
We have developed a new method incorporating some of the ideas of DMRG and the
transfer matrix method successfully used in the non-interacting case
\cite{rf:3a,rf:3b}.  Section~\ref{chap:method} describes the method and
section~\ref{chap:single} discusses the application to a model of spinless
fermions.

\section{The New Method}
\label{chap:method}
 Like DMRG our approach is based on a tight-binding method and has a similar potential usefulness.  It can be readily applied to
 describe any 1D or quasi-1D system, provided interactions are
 nearest-neighbor. 
\subsection{The Hamiltonian}
As this is a many-body problem it is natural to work within the second quantization formalism of quantum mechanics.
This allows the Hamiltonian to be written in terms of particle creation $\crt{i}$ and annihilation
 $\ann{i}$ operators for site $i$:
\begin{eqnarray}
\label{eq:ham}
\hat{H}&=&\sum_i\varepsilon_i\crt{i}\ann{i}
+V\sum_i(\crt{i}\ann{i+1}+\crt{i+1}\ann{i})\nonumber\\
&&+U\sum_i(\crt{i}\ann{i})(\crt{i+1}\ann{i+1})
-\mu\sum_i\crt{i}\ann{i} .
\end{eqnarray}
The first two terms constitute the standard Anderson model\cite{anderson} used widely
 in the study of disorder induced localization.  The additional $U$ term
 represents the nearest-neighbor interaction.
 If neighboring sites are occupied then the two particles experience a repulsive force (as for electrons) or possibly attractive
 force of strength $U$.  Setting $U=0$ reverts the system to the many
 independent-body situation (\ie the non-interacting case).

The final term represents the chemical potential $\mu$.  It is necessary as this method works within the
 grand canonical scheme in which a range of particle numbers will be considered.  The value of the parameter $\mu$ corresponds to
the energy required to add a particle to the system, and thus controls the
particle density of the system ground state.  As with most numerical studies of Anderson localization zero temperature will be assumed.

\subsection{The Recursive Method}
Our method tackles the problem of the exponentially growing Hilbert space by reducing the number of basis states,
 restricting the focus to the ground state.  This works in conjunction with a recursive procedure that
 extends the chain by successively adding new sites.  Open boundary conditions must be used.

\begin{figure}[h]
\centering
\includegraphics[width=\mmin{6cm}{\linewidth}]{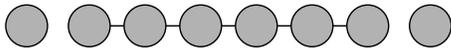}
\caption[Schematic picture of constructing the single chain]
{The recursive procedure adds new sites to both ends of a 1D chain at each iteration.}
\label{chain}
\end{figure}

For each iteration:
\begin{itemize}
\item A site is added to each end of the chain (fig.~\ref{chain}) and basis states are constructed.  At first sight it may appear simpler to add a new site to one end only.  However, it turns out that for the purposes of measuring the degree of localization it is much more natural to add sites to both ends of the chain in the same iteration (section \ref{sec:local}).

\item For each particle number with remaining basis states a Hamiltonian matrix is found.  After the matrix elements
 have been calculated, the Hamiltonian is solved and the desired quantities are extracted.

\item Finally, a proportion of the resulting eigenstates are thrown away according to some criterion.  The remaining states are used to form the basis at the next iteration - a chain with two more sites.
\end{itemize}

There is no fundamental reason why this process cannot be repeated to very large chain lengths.  
The following sections detail the mathematics of this procedure.

\subsubsection{Expressing the Basis States}
\label{sec:expbasis}
For a one-dimensional chain, with $L$ sites and one electron per site, there are $2^L$ basis states.  In order to reduce
the Hilbert space, this
 new method relies on the fact that it is possible to express the states for a
 chain of length $L$ in terms of the energy eigenstates $\ket{\Phi^{L-2}}$ of the chain of length $L-2$ (\ie the
 same chain without the two end sites).  Formally, the $L$ site Hilbert space is a product of the Hilbert space for the
 $L-2$ site chain with the vector space associated with the two new sites.

The eigenstates $\ket{\Phi^{L-2}}$ can be used as an orthogonal basis for the inner Hilbert space because they are
 eigenvectors of the previous iteration Hamiltonian such that
\begin{equation}
\label{eq:eigen}
\hat{H}^{L-2} \ket{\Phi^{L-2}_i} = E_i \ket{\Phi^{L-2}_i}.
\end{equation}

 The outer Hilbert space, associated with the two new sites, is spanned by four basis states.  This is readily
 seen by considering particle number occupancy representation:
 $\ket{0\cdots 0}$, $\ket{1\cdots 0}$, $\ket{0\cdots 1}$ and $\ket{1 \cdots 1}$.  Thus for every
 eigenstate $\ket{\Phi^{L-2}_i}$ of the $L-2$ site chain, there are four corresponding basis states for the $L$
 chain: $\ket{0 \Phi^{L-2}_i 0}$, $\ket{1\Phi^{L-2}_i 0}$, $\ket{0\Phi^{L-2}_i 1}$ and $\ket{1\Phi^{L-2}_i 1}$.

 Consequently, a general state for the $L$ chain with $N$
 electrons, $\ket{\Psi_n^{L,N}}$, may be written as a linear combination
 of basis states in the following manner:
\begin{eqnarray}
\label{eq:expand}
\lefteqn{\ket{\Psi_n^{L,N}}=\phantom{+}\sum_ia_{ni}\ket{0\Phi_i^{{L-2},N}0}}\nonumber\\
&+&\sum_j \left\{ b_{nj}\ket{1\Phi_j^{{L-2},N-1}0}
+c_{nj}\ket{0\Phi_j^{{L-2},N-1}1} \right\} \nonumber \\
&+&\sum_kd_{nk}\ket{1\Phi_k^{{L-2},N-2}1} .
\end{eqnarray}
In fact, as this equation indicates, it is only necessary to consider the subset of $\ket{\Phi^{L-2} }$ states which,
when combined with two new end sites, have a total of $N$ electrons.  The reason is that particle number is a good quantum number for this Hamiltonian. 

\subsubsection{Calculating the Hamiltonian Matrix}
\label{sec:ham}
Thus basis states can be grouped according to particle number and a separate Hamiltonian can be calculated for each.
This can only be accomplished by first expanding each of the four types of basis states for $N$ particles and ${L-2}$
 sites back a further generation, in terms of the previous iteration $\ket{\Phi^{L-4}}$:
\begin{eqnarray}
\ket{m\,\Phi_i^{N}\, n}
&=&\phantom{+}\sum_{p}a^{mn}_{ip}\ket{m\,0\,\Phi_p^{N\phantom{+1}}\,0\,n}\nonumber\\
&&+\sum_q b^{nm}_{iq}\ket{m\,1\,\Phi_q^{N-1}\,0\,n}\nonumber\\
&&+\sum_r c^{mn}_{ir}\ket{m\,0\,\Phi_r^{N-1}\,1\,n}\nonumber\\
&&+\sum_s d^{mn}_{is}\ket{m\,1\,\Phi_s^{N-2}\,1\,n}
\end{eqnarray}
where $m,n = 0,1$ and the $L-4$ superscripts have been dropped for the sake of clarity.  It is now possible to cast the Hamiltonian
 in a corresponding form.  This involves some tedious algebra.  However,
 bearing in mind that the states for different $N$ are orthogonal as are the
 sets of eigenstates $\Phi_p^{L-2}$ and $\Phi_q^{L-4}$ of the Hamiltonian at the 2 previous iterations, the final Hamiltonian may be written as the following $4\times4$ block form:
\begin{widetext}
\centering
\begin{equation*}
\label{eq:hamiltonian}
\begin{array}{|c||c|c|c|c|}
\hline
&&&&\\
&\ket{0\Phi_i^{N}0}&\ket{1\Phi_j^{N-1}0}&\ket{0\Phi_j^{N-1}1}&\ket{1\Phi_k^{N-2}1}\\
&&&&\\
\hline\hline
&&&&\\
\bra{0\Phi_{i'}^{N}0}
&E_i\delta_{i'i}
&\begin{array}{c} V\sum\limits_qb^{00}_{i'q}a^{10}_{jq} \\
+V\sum\limits_rd^{00}_{i'r}c^{10}_{jr} \end{array}
&\;\begin{array}{c} V\sum\limits_qc^{00}_{i'q}a^{01}_{jq} \\
+V\sum\limits_rd^{00}_{i'r}b^{01}_{jr} \end{array}\;
&0\\
&&&&\\
\hline
&&&&\\
\bra{1\Phi_{j'}^{N-1}0}
&\;\begin{array}{c} V\sum\limits_qa^{10}_{j'q}b^{00}_{iq}\\
+V\sum\limits_rc^{10}_{j'r}d^{00}_{ir} \end{array}\;
&\begin{array}{c} (E_j+\varepsilon_{1}-\mu)\delta_{j'j}\\
+U\sum\limits_rb^{10}_{j'r}b^{10}_{jr}\\
+U\sum\limits_sd^{10}_{j's}d^{10}_{js} \end{array}
&0
&\begin{array}{c} V\sum\limits_rc^{10}_{j'r}a^{11}_{kr}\\
+V\sum\limits_sd^{10}_{j's}b^{11}_{ks} \end{array}\\
&&&&\\
\hline
&&&&\\
\bra{0\Phi_{j'}^{N-1}1}
&\begin{array}{c} V\sum\limits_qa^{01}_{j'q}c^{00}_{iq}\\
+V\sum\limits_rb^{01}_{j'r}d^{00}_{ir} \end{array}
&0
&\;\begin{array}{c}(E_j+\varepsilon_{L}-\mu)\delta_{j'j}\\
+U\sum\limits_rc^{01}_{j'r}c^{01}_{jr}\\
+U\sum\limits_sd^{01}_{j's}d^{01}_{js} \end{array}\;
&\begin{array}{c}V\sum\limits_rb^{01}_{j'r}a^{11}_{kr}\\
+V\sum\limits_sd^{01}_{j's}c^{11}_{ks} \end{array}\\
&&&&\\
\hline
&&&&\\
\;\bra{1\Phi_{k'}^{N-2}1}\;
&0
&\begin{array}{c} V\sum\limits_ra^{11}_{k'r}c^{10}_{jr}\\
+V\sum\limits_sb^{11}_{k's}d^{10}_{js} \end{array}
&\begin{array}{c} V\sum\limits_ra^{11}_{k'r}b^{01}_{jr}\\
+V\sum\limits_sc^{11}_{k's}d^{01}_{js} \end{array}
&\;\begin{array}{c} (E_k+\varepsilon_{1}+\varepsilon_{L}-2\mu)\delta_{k'k}\\
+U\sum\limits_sb^{11}_{k's}b^{11}_{ks}+U\sum\limits_sc^{11}_{k's}c^{11}_{ks}\\
+2U\sum\limits_td^{11}_{k't}d^{11}_{kt} \end{array}\;\\
&&&&\\
\hline
\end{array}
\end{equation*}
\end{widetext}
For each particle number with a set of basis states, this block matrix can be used to generate the elements of
 the full matrix.  A ground state can be calculated for each of these particle numbers.  The ground state lowest
 in energy is the system ground state (this is how $\mu$ controls the ground state particle density).

\subsubsection{Reducing the Number of Basis States}
\begin{figure}[htb]
\centering
\includegraphics[width=\linewidth]{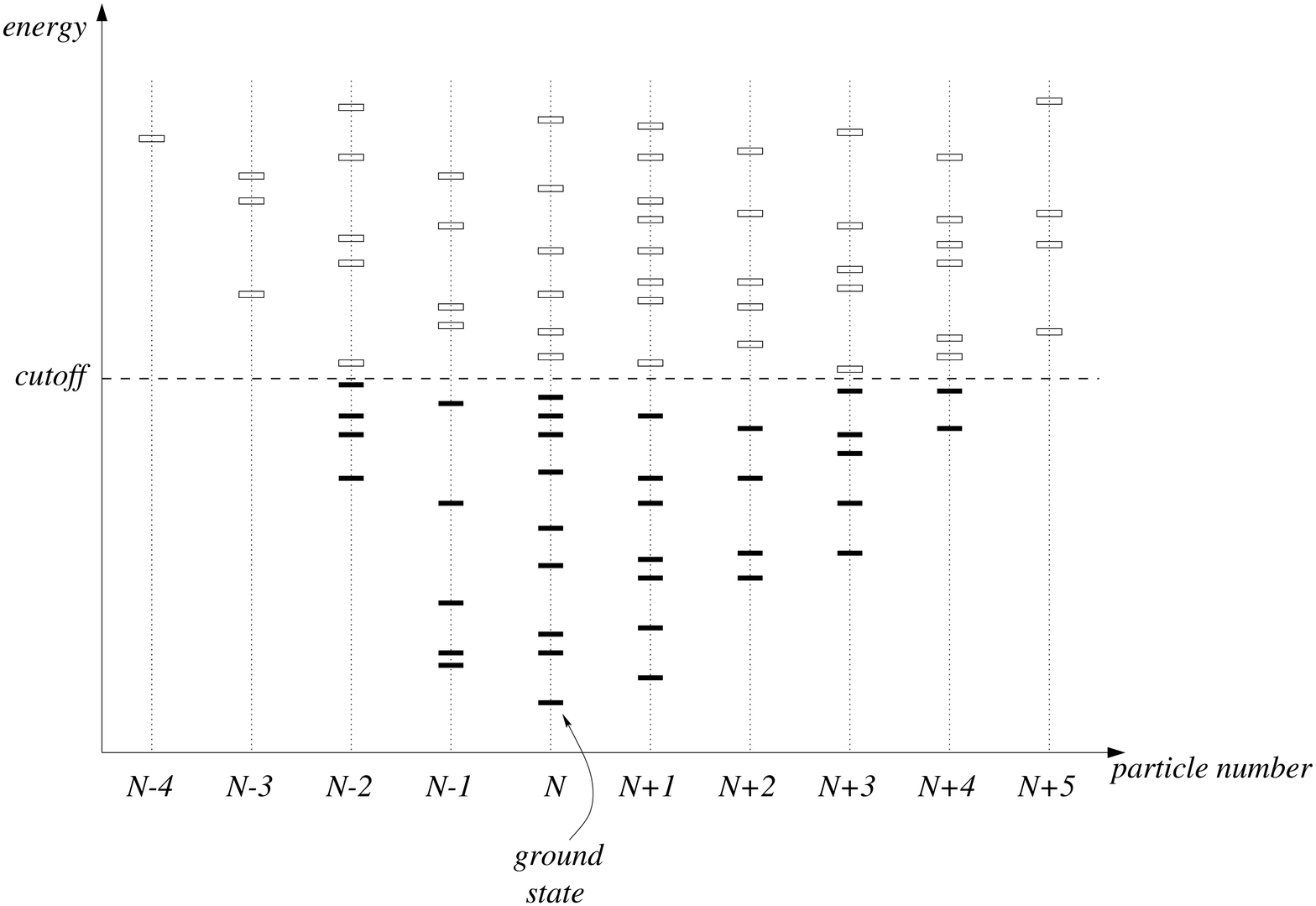}
\caption[Diagram of the procedure for systematically discarding states]
{This diagram illustrates the simple energy cutoff procedure for reducing the Hilbert space.
 It shows an example of basis states scattered according to energy and particle number.  Firstly, the ground state is located.
 Then the excited states with the same occupation number are counted in order to set the cutoff energy as the midpoint
between the $M$th and $(M+1)$th state (in this case $M=9$).  All states below the cutoff are kept and the rest are
discarded.}
\label{cutoff}
\end{figure}
The purpose of reformulating the basis states and in turn the Hamiltonian in
this manner is to enable an approximation to be introduced which keeps the
dimension of the Hilbert space roughly constant as sites are added. During
each iteration a proportion of the basis states must be thrown away according
to some systematic method. This is necessary to keep the calculation to a
computationally manageable size.  Within the tight-binding framework it is
the only approximation in our method.

There are several possible schemes which could be used.  A criterion is
required that produces the smallest error on the next iteration ground state
as it is the properties of the ground state which are of interest.  Na{\"\i}vely, the lowest energy states could be kept.  More sophisticated approaches would determine which states make the largest contribution to the next iteration ground state.  Whichever method is adopted, some justification will be required as will the determination of the limits of its accuracy.

The simplest method to implement is to throw away the states of highest energy, so this will be adopted initially.  The diagonalization routines automatically sort eigenstates according to their eigenenergies making the procedure relatively straightforward.  Thus during each step, after diagonalization but before extending the chain, the highest energy states are discarded.  This is achieved by setting an energy cutoff halfway between the $M$th and $(M+1)$th eigenvalue with the same occupation number as the system ground state.  This is demonstrated in fig.~\ref{cutoff}.  Then all states with energy higher than the cutoff are removed.  The value of $M$ can be changed to control the accuracy where higher accuracy of course entails larger processing time and memory requirements.

\subsection{Measuring the Localization Length}
\label{sec:local}
The aim of this new method is to understand the effect of varying system
parameters, in particular the interaction strength $U$, on electron
localization.  For non-interacting systems and assuming the wavefunction has an exponentially decaying envelope, the degree of localization can be characterized by the localization length $\lambda$.  This quantity is related to the transport and conductivity of the system.

In the non-interacting case a number of methods have been implemented for
extracting the localization length\cite{angus}.  However, most methods of
gauging localization cannot be simply carried across into the many-body case.
The sensitivity to boundary conditions (BCs)\cite{schmit,angus} does not suffer from this problem.
 
This method works by bringing the ends of the chain together to form a ring, which is equivalent to changing the boundary conditions from open to periodic.  In fact, in so doing it is possible to introduce a complex phase factor.  The essence of the method is to calculate the change in ground state energy as the boundary conditions are twisted, \ie as the phase factor is varied.  For a localized state very little change should  be observed as the wave amplitude has decayed to zero.  However, an extended state should experience a substantial change since the wave amplitude has not decayed off.

The simplest way to implement this approach is to use the \textit{phase sensitivity} $D$.
This is defined as the difference in ground state energy $E_0$ between the
system with periodic BCs and the system with anti-periodic BCs:
\begin{equation}
D=E_0(\mbox{periodic})-E_0(\mbox{anti-periodic}) .
\end{equation}
The length dependence of the phase sensitivity can be used to define $\lambda$:
\begin{equation}
D \propto e^{-\frac{L}{\lambda}}.
\end{equation}
As the new recursive method works by extending a chain with open boundary conditions, it was necessary to implement the phase sensitivity perturbatively.  This was done both as an analytical perturbation and as a numerical perturbation.  The former actually reduces to calculating certain elements of the density matrix.

\subsubsection{Analytical Perturbation} \label{sec:analpert} 
To implement the phase sensitivity as an analytical perturbation, consider the first order energy shift of the ground state $\ket{\Phi_0}$
\begin{equation}
\delta E_0=\bra{ \Phi_0 } \delta \hat{H} \ket{ \Phi_0 },
\end{equation}
where $\delta \hat{H}=\pm V(\crt{1}\ann{L}+\crt{L}\ann{1})+U(\crt{1}\ann{1})(\crt{L}\ann{L})$ contains the hopping and interaction terms now connecting the two ends of the chain.  Calculating the effect of $\delta \hat{H}$ motivates adding sites simultaneously to both ends of the chain.  Given
that $\ket{\Phi_0}$ is a linear combination of basis states (see eqn.~\ref{eq:expand}), the effect of $\delta \hat{H}$ on $\ket{\Phi_0}$ is
\begin{eqnarray}
\lefteqn{\delta \hat{H} \ket{\Phi_0^{L,N}} = \qquad}\nonumber\\
&\pm& V(-1)^{s_{L-1}} \sum_j \left\{
b_{0j}\ket{0\Phi_j^{{L-2},N-1}1}\right.\nonumber\\
&&\qquad\qquad +\left.c_{0j}\ket{1\Phi_j^{{L-2},N-1}0}
\right\} \nonumber \\
&+&U \sum_kd_{0k}\ket{1\Phi_k^{{L-2},N-2}1}.
\label{eq:deltaH}
\end{eqnarray}
where $(-1)^{s_{L-1}}$ is a phase factor due to electron anti-symmetry
arising out of the occupancy of the inner $L-2$ sites. Substituting this into the expression for $\delta E_0$ gives
\begin{equation}
\delta E_0=\pm 2V(-1)^{s_{L-1}} \sum_j b_{0j}c_{0j} + U \sum_k(d_{0k})^2.
\end{equation}

The phase sensitivity is the difference between periodic and anti-periodic energy shifts, thus the last
 interaction term cancels yielding
\begin{equation}
\label{eq:ps}
D=4V (-1)^{s_{L-1}} \sum_j b_{0j}c_{0j}.
\end{equation}
The information required to calculate the factor $(-1)^{s_{L-1}}$ is unavailable.  Fortunately the degree of localization can be calculated by using $D^2$ instead.  Alternatively, this problem can also be avoided by choosing a different order for applying creation operators.

The ``scalar product'' quantity in the expression for $D$ corresponds to
calculating the off-diagonal element of the reduced density matrix, 
\begin{equation}
\label{eq:rdm}
\rho_{\bra{1\cdots 0},\ket{0\cdots 1}}=\rho_{\bra{0\cdots 1},\ket{1\cdots
0}}=\sum_jb_{0j}c_{0j}.
\end{equation}
This is intuitively unsurprising because the quantity of interest is the probability that given an electron is placed at one end of the chain, \textit{an} electron comes out the other end.  Furthermore, only information about the ends of the chain is available (and required), so a \textit{reduced} density matrix is used that sums over the redundant middle part of the chain.

\subsubsection{Numerical Perturbation}
\label{sec:phase}
The second way the phase sensitivity to boundary conditions was implemented uses a numerical perturbation.  That is, the calculation proceeds as normal with open boundary conditions.  At each step, working with the normal basis states two additional Hamiltonians are formed corresponding to periodic and anti-periodic BCs. These are then solved and the ground state for each type of BC is found.  The phase sensitivity is then easily calculated.  The basis states generated by the \textit{open} BC Hamiltonian are kept as normal, but the
 states from the other two BCs are discarded.

Using this approach means performing extra diagonalizations, so it is computationally time consuming.  Hence it was used to numerically
 verify the legitimacy of the analytical perturbation (\ie the off-diagonal element of the reduced density matrix).

The Hamiltonian for (anti-)periodic BCs is identical to the open BCs Hamiltonian (\ref{eq:hamiltonian})
but with a few extra terms.  These additional terms are those calculated in
the analytical perturbation for each basis state (\ref{eq:deltaH}) and appear in the diagonal matrix elements of the Hamiltonians.

\subsection{Model Properties}
The Hamiltonian models used with this new method possess some shared properties.  These will be defined in this section,
again in terms of the single chain model, and suggest some consistency tests which can be used to provide limited justification for the method.

\subsubsection{Definition of Energy Gaps}\label{sect:gaps}
Several definitions of the energy gap exist in many-body systems.  The
numerical method under development works within the grand canonical
(GC) framework so eigenvalues are grand canonical energies.  Thus to convert
to canonical (C) energies the following relation must be used
\begin{equation}
E^{GC}(N)=E^C(N)-N\mu.
\end{equation}
Various energy gaps are defined in (\ref{eq:gaps})
\begin{subequations}
\label{eq:gaps}
\begin{eqnarray}
\Delta E_{\mbox{\scriptsize ph}} &=& E_1(N)-E_0(N) \label{eq:gap:1}\\
\Delta E_+ &=& E_0(N+1)-E_0(N) \label{eq:gap:2}\\
\Delta E_- &=& E_0(N)-E_0(N-1) \label{eq:gap:3}\\
\Delta E &=& E_0(N+1)+E_0(N-1)-2E_0(N) .\label{eq:gap:4}
\end{eqnarray}
\end{subequations}
where $\Delta E_{\mbox{\scriptsize ph}}$ (\ref{eq:gap:1}) is the difference
between the ground state and the 1st excited state, $\Delta E_+$ and $\Delta
E_-$ (\ref{eq:gap:2} \& \ref{eq:gap:3}) are the energies to add or remove and
electron from the system, and $\Delta E$ is the difference of $\Delta E_+$ and
$\Delta E_-$ (\ref{eq:gap:4}).
When transferring to canonical energies, $\Delta E_+$ and
 $\Delta E_-$ will shift by a constant $\mp \mu$.  The non-interacting limit yields some predictions which simulations should reproduce.

\subsubsection{Length Dependence}
In the single body case there is one state per site and the bandwidth is constant so the density of states
 is proportional to the number of sites $L$.  This means on average $\Delta E_{ph} \propto \frac{1}{L}$ at the
  same position within the band.  The canonical and grand canonical versions are identical as $\Delta E_{ph}$
 is a difference between states with identical particle number.

The \textit{canonical} energy required to add a particle should be a little larger than the chemical potential.
In the ground state, the highest occupied state will be the first state below the chemical potential.  In order to add a particle the energy of the first state above the chemical potential is
required.  Thus the canonical energy is the chemical potential plus a contribution with a length dependence of
 $\frac{1}{L}$.  The maximum
of this contribution is $\Delta E_{ph}$ corresponding to the case in which, before adding the electron, the highest occupied
state energy was precisely the chemical potential energy.  When transferring into the grand canonical scheme $\Delta E_{+}$
reduces to the $\propto \frac{1}{L}$ contribution.  In the middle of the band, where $\mu=0$, the canonical and grand
canonical energies are identical.

\subsubsection{Consistency Test}
\begin{figure}[htb]
\centering
\includegraphics[width=\mmin{4cm}{\linewidth}]{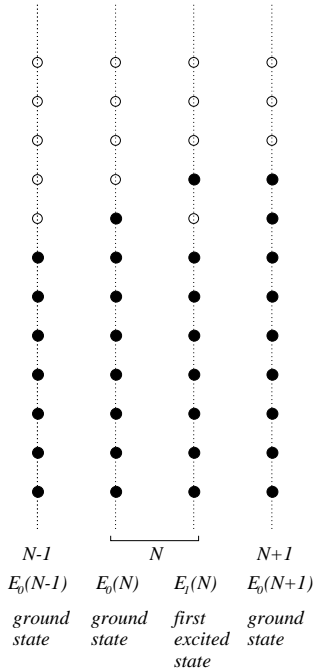}
\caption[Diagram illustrating the consistency test relation between ground state energies]
{Illustrates the consistency test relation between ground state energies.  The filled dots are
 occupied single particle states.}
\label{sctest1}
\end{figure}
The energy gap definitions also give a consistency test that works in the presence of disorder, but unfortunately 
 does not work in the presence of Coulomb interactions.  Consider the energy change
 when one electron is added to the ground state.  It can readily be seen that a relation can be found for
 the difference between ground state energy of the
system $E_0(N)$ and the ground state with an additional particle $E_0(N+1)$
\begin{equation}
E_0(N+1)-E_0(N)=E_1(N)-E_0(N-1),
\end{equation}
where $E_1(N)$ is the first excited state with the same particle number as the system ground state.  
Figure~\ref{sctest1} shows the four states referred to in the relation.
The right-hand side can be substituted into the definition of $\Delta E$, then after canceling $\Delta E$
reduces to $\Delta E_{ph}$.  Therefore the above relation is equivalent to
\begin{equation}
\label{eq:gaptest}
\Delta E = \Delta E_{ph}.
\end{equation}

The accuracy to which these equivalence relations will be satisfied
depends on the number of basis states retained in each generation.  If all states are kept this should be an exact formula.

\subsubsection{Particle-Hole Symmetry}
When a band is nearly full it is often more instructive to consider the system in terms of holes rather than particles.
 Even in the case with disorder and interactions there should be a symmetry between the behavior of holes in the top
 of the band and the behavior of electrons in the bottom of the band.  The symmetry relation between the two
 can be found by rewriting the Hamiltonian in terms of holes and then comparing with the original electron Hamiltonian.

The Hamiltonian for the system given in terms of electron occupancy is (for open boundary conditions):
\begin{eqnarray}
\hat{H} &=& \sum_{i=1}^L \crt{i}\ann{i}\varepsilon_i
+V\sum_{i=1}^{L-1}(\crt{i}\ann{i+1}+\crt{i+1}\ann{i})\nonumber\\
&&+U\sum_{i=1}^{L-1}(\crt{i}\ann{i})(\crt{i+1}\ann{i+1})
-\mu \sum_{i=1}^L\crt{i}\ann{i}.
\end{eqnarray}
To rewrite this for holes it is necessary to define suitable creation and annihilation operators for holes, $\hcrt{i}$
 and $\hann{i}$ respectively.  The equivalence relations to particle operators are:
 $\hcrt{i}=\ann{i}(-1)^i$ and $\hann{i}=\crt{i}(-1)^i$.  Substituting these into the electron
 Hamiltonian and rearranging gives a similar form of Hamiltonian to the original
\begin{eqnarray}
\hat{H}&=&-\sum_{i=1}^L\hcrt{i}\hann{i}\varepsilon_i
+V\sum_{i=1}^{L-1}(\hcrt{i}\hann{i+1}+\hcrt{i+1}\hann{i})\nonumber\\
&&+U\sum_{i=1}^{L-1}(\hcrt{i}\hann{i})(\hcrt{i+1}\hann{i+1})
+(\mu-2U)\sum_{i=1}^L\hcrt{i}\hann{i}\nonumber\\
&&+\sum_{i=1}^L\varepsilon_i+(L-1)U-L\mu+U\hcrt{1}\hann{1}+U\hcrt{L}\hann{L} .
\end{eqnarray}
The last two terms arise from using open boundary conditions.  In order to correct for these terms, from now on
the original Hamiltonian will be adjusted by adding
$+\frac{U}{2}\crt{1}\ann{1}+\frac{U}{2}\crt{L}\ann{L}$.  In the algorithm
 these terms are readily introduced but must only be applied to the end sites, \ie they must be
 removed when extending the chain length.  Once added, the equivalent Hamiltonian in terms of holes then becomes
\begin{eqnarray}
\hat{H}&=&-\sum_{i=1}^L\hcrt{i}\hann{i}\varepsilon_i
+V\sum_{i=1}^{L-1}(\hcrt{i}\hann{i+1}+\hcrt{i+1}\hann{i})\nonumber\\
&&+U\sum_{i=1}^{L-1}(\hcrt{i}\hann{i})(\hcrt{i+1}\hann{i+1}) \nonumber \\
&&+\frac{U}{2}\hcrt{1}\hann{1}+\frac{U}{2}\hcrt{L}\hann{L} \nonumber\\
&&+(\mu-2U)\sum_{i=1}^L\hcrt{i}\hann{i}+\sum_{i=1}^L\varepsilon_i+L(U-\mu).
\end{eqnarray}
This symmetry means that an exact correspondence of eigenvectors is expected to be observed at the top and bottom
of the band such that
\begin{eqnarray}
\label{eq:ehtest}
\lefteqn{\hat{H}(\varepsilon_i=\varepsilon_i',\mu=-\mu',U=U') =
\qquad}\nonumber\\
&&\hat{H}(\varepsilon_i=-\varepsilon_i',\mu=\mu'+2U',U=U')\nonumber\\
&&+\sum_{i=1}^L\varepsilon_i'+L(U'+\mu'),
\end{eqnarray}
with a shift of eigenenergies according to the last two terms. In fact,
the $+\sum_{i=1}^L\varepsilon_i'+L(U'+\mu')$ terms represent the energy of the full electron band which can be easily
seen from the electron Hamiltonian.  The significance is that the hole picture starts from the
 top of the band and works downward.

Moreover, because the eigenvectors are identical, the reduced density matrix method for calculating localization
 length will give exactly the same results either top or bottom of the band.  This is because
  $\rho_{\bra{0\cdots 1},\ket{1\cdots 0}} \equiv \rho_{\bra{1\cdots 0},\ket{0\cdots 1}}$ from eqn.~\ref{eq:rdm}, and so can be used as a consistency test.

This symmetry may be extended from a single sample to the average case by noting that the probability distribution
 for $\varepsilon_i$ is a symmetric function about the origin.  This has two implications. Firstly, the
$\sum_{i=1}^L\varepsilon_i$ term will tend to zero as the number of sites is increased.  And secondly, the sign
 of the $-\sum_{i=1}^L\hcrt{i}\hann{i}\varepsilon_i$ term can be flipped.  This simplifies the equivalence
 relation to
\begin{eqnarray}
\label{eq:ehtest2}
\lefteqn{\hat{H}(\mu=-\mu',U=U') = \qquad}\nonumber\\
&&\hat{H}(\mu=\mu'+2U',U=U')+L(U'+\mu').
\end{eqnarray}
for an ensemble average.

The particle-hole symmetry also gives the condition for half-filling: $\mu'=U'$.  This can be seen
 by setting $-\mu'=\mu'+2U'$, so that in both eqns.~\ref{eq:ehtest} and \ref{eq:ehtest2} the same value
 for $\mu$ is specified.

\subsection{Computational Implementation}
\begin{figure}[th]
\centering
\includegraphics[width=\mmin{6cm}{\linewidth}]{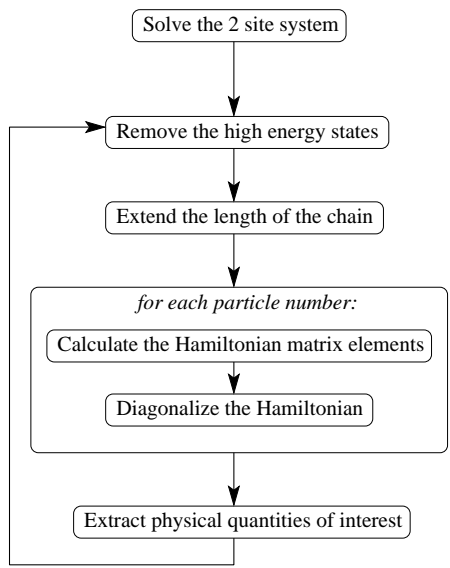}
\caption[Flow diagram of the central algorithm of the new method]
{A flow diagram showing the central procedure of the algorithm.  This whole procedure
 is repeated many times with different disorder realizations.}
\label{flow}
\end{figure}
The simulation was written in C++, making use of the object orientated facilities.  Because states need to be added
and removed from a set of states it was natural to use linked lists, with each link holding data for one state and the entire
list representing a set of basis states or eigenstates.  This results in making the code for the central algorithm less
cumbersome.  The structure of the central algorithm is shown in fig.~\ref{flow}.

To extract sensible data localization quantities must be averaged over many systems.  Conventional practice is to
 perform a geometric average which is achieved by averaging the logarithm of the phase sensitivity.  Then least-square
 fits were carried out to extract the localization length over a minimum of 10 sites.  In addition other quantities
  were recorded such as the particle density, ground state energy and energy gaps.

\subsubsection{Removing States}
\begin{figure}[h]
\centering
\includegraphics[width=\linewidth]{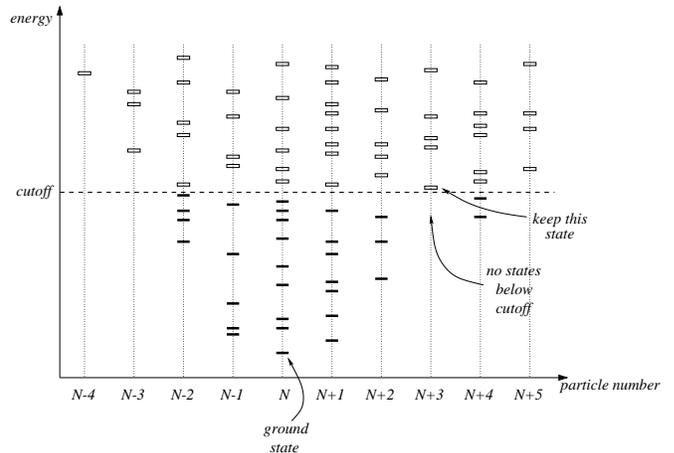}
\caption[Diagram demonstrating the necessary modification of the discarding states procedure]
{This diagram illustrates how the simple energy cutoff scheme may occasionally cause
 difficulties.  All states with $N+3$ particles will be removed, however some states with $N+4$ particles
  will be retained.  To avoid considerable overhead in coding, the lowest energy state for $N+3$ will be kept.}
\label{cutoff2}
\end{figure}
 One minor modification to the above procedure was made.  It is simplest to code this new numerical method assuming
that states exist for every particle number between two limits (in a range roughly centered around the ground state particle
density).  For example in fig.~\ref{cutoff}, after the Hilbert space reduction, each
 value of $N$ between $N-2$ and $N+4$ inclusive have states remaining.  With the reduction scheme outlined above a
 scenario occasionally arises whereby this assumption would be rendered
invalid.  Figure~\ref{cutoff2} demonstrates
 that it is possible for one value of $N$, in the fig.~$N+3$, to have all states removed yet neighboring particle
 numbers to have states left.  To cater for this rare event the code would become unnecessarily complicated.  The problem can be easily
 overcome in such cases by retaining the lowest state even though it is above the energy cutoff.

 In addition to this, the ground states corresponding to $N+1$ and $N-1$ particle numbers were always retained at each iteration.
This ensures the various energy gaps can be calculated even in the rare event when the $N-1$ or $N+1$ ground state lies
 above the cutoff energy.

\section{The Single Chain Model}
\label{chap:single}

This method was first applied to the Hamiltonian (\ref{eq:ham}) plus the two correction terms to ensure particle-hole symmetry:
\begin{eqnarray}
\hat{H}&=&\sum_{i=1}^L\crt{i}\ann{i}\varepsilon_i
+V\sum_{i=1}^{L-1}(\crt{i}\ann{i+1}+\crt{i+1}\ann{i})\nonumber\\
&&+U\sum_{i=1}^{L-1}(\crt{i}\ann{i})(\crt{i+1}\ann{i+1})
+\frac{U}{2}\crt{1}\ann{1}+\frac{U}{2}\crt{L}\ann{L}\nonumber \\
&&-\mu\sum_{i=1}^L\crt{i}\ann{i}.
\end{eqnarray}
This is the conventional 1D Anderson model with nearest neighbor interactions.  
In this paper
hopping will always be set to $V=1$ (hence defining the energy scale for both $U$ and $W$).  The following two subsections outline
some useful features already known about this model in two limits: no interactions ($U=0$) and no disorder ($W=0$).

\subsection{Non-Interacting Behavior}
A single chain with one orbital per site has one band centered on zero with
bandwidth $4V$.  Including disorder will blur the edges, effectively widening
the band.  The localization properties of a one-dimensional
 non-interacting chain are well established.  For any amount of disorder
all eigenstates are localized.  The dependence of localization length on disorder is usually
 quoted as \cite{angus}
\begin{equation}
\lambda^{-1} = \frac{W^2}{24(4V^2-\mu^2)}.
\label{eq:niresult}
\end{equation}
This is only valid for small disorder.  Note that the localization length
diverges in the clean limit.  Therefore, an important test for the new
recursive method is to reproduce this behavior.  However, care is required in
making the correct comparison: how does this dependence carry across from the
single-particle case to the many-particle case?  A simple test program was constructed using exact diagonalization to calculate the phase sensitivity to boundary conditions for both the single state in the centerer of the band and the phase sensitivity for the grand canonical energy.  Good agreement was found.

\subsection{Clean Phase Space}
The second limit to be outlined is the zero disorder phase space.  This is understood because without randomness
 the present model can be mapped to a XXZ spin chain model and solved exactly
 for half-filling \cite{pang,yang1,yang2,yang3}.
 
A program was constructed to perform exact diagonalizations on short chains in order to compare results.  This was
necessary because
 the method under development doesn't use the particle occupation basis.
 The computational limit is about 10 sites,
but nevertheless gives valuable insight into the nature of the ground state for different regions of phase space.

At half-filling there are two limiting forms of the ground state with a crossover regime.  For large repulsive
 interactions a charge density
wave (CDW) is observed (\ie alternate sites are occupied).  For attractive interactions, electrons tend to
cluster together.  For open boundary conditions (with the correction for particle-hole symmetry) the transition region
 occurs between $U=-0.8$ and $U=-2$.
\begin{figure}[ht]
\centering
\includegraphics[width=\mmin{10cm}{\linewidth},clip]{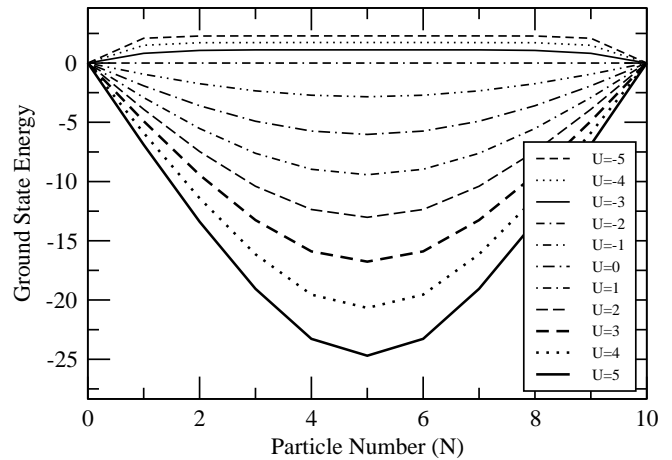}
\caption[Demonstration of phase separation for a short single chain]
{Results from a (clean) short chain of 10 sites demonstrating phase separation for $U<-2$.  For each
 particle number the ground state energy is plotted.  The chemical potential
 is set to give half-filling as
 the overall ground state (\ie $\mu=U$). Plotted energies are grand canonical.}
\label{psep}
\end{figure}

For attractive interaction below $U=-2$, it is impossible to maintain
half-filling within the grand canonical scheme. The ground state is a completely empty or completely full band (\ie it is unstable to phase separation) as can be seen in figure~\ref{psep}.  In fact, as the $U=-2$ limit is reached from above, the ground state energy tends toward being independent of particle number $N$.
      
In contrast, for increasing repulsive interactions, above $U=2$ a charge gap opens up \cite{pang,bouz}.  In other words, the CDW above this point
corresponds to a Mott insulator state.  For short chains it is not possible to
pinpoint where this gap begins.

\subsection{Previous Work}
\label{sec:ropw}
Perhaps the most substantial work is that of Giamarchi and Schulz \cite{giam1}, although they acknowledge an earlier paper \cite{apel}.
For the present one-dimensional model repulsive interactions increase localization because the CDW is pinned by the disorder.  In contrast, attractive
interactions decrease localization.  In fact, a delocalized phase is predicted for sufficiently attractive interactions.
  Giamarchi and Schulz develop a $k$-space Renormalization
Group approach to study the transition.  The existence of this transition is ascribed to competition between disorder and superconducting fluctuations \cite{giam1}.

Numerical work has sought to verify these predictions and in particular to map
out the delocalized regime.  One paper\cite{bouz} performs exact diagonalizations on small systems (up to 22 sites).  The results are consistent with the expected delocalized phase, although because the chain length is so small they cannot exclude the possibility that the localization length is very large.

The first DMRG study \cite{pang} focused on the effect of disorder on the Mott
state.  The authors conclude that even weak disorder destroys the charge gap
and long-range order associated with the CDW state (although the nature of elementary excitations remain unchanged).

The most extensive work has been conducted by Schmitteckert \etal applying DMRG to both the
 interacting Anderson model and to the
 related problem of persistent currents in mesoscopic rings \cite{schmit3,schmit,schmit4,schmit2}.
The first study examining Anderson localization \cite{schmit} was on chains extending up to 60 lattice sites.
 The degree of localization was measured by the phase sensitivity to boundary conditions.  Two regimes were found: a
 localized phase, $U>-1$, and delocalized regime, $U<-1$, consistent with work already mentioned.
   In fact, it was found repulsive interactions increase localization.  After considerable
 numerical effort a phase diagram was produced showing where the two regions
 lie in disorder-interaction space.   The authors believe the earlier attempt \cite{bouz} based on an RG procedure over estimates the delocalized regime by a factor of 4. Other authors, R{\"o}mer\cite{romer1} and Schuster \etal\cite{romer3}, have mapped out an extended regime for the same model but with the Aubry-Andr\'{e} quasi-periodic potential.  However its shape in disorder-interaction phase space takes on a different form.

In two more recent papers \cite{schmit4,schmit2} Schmitteckert \etal showed that important physics is washed out in the process of ensemble averaging used in ascertaining numerical data.  They examined the chaotic region between a chain characterized as an Anderson insulator and characterized as a Mott insulator corresponding to the strongly disordered and the strongly correlated limits respectively.

Schmitteckert \etal showed that in this transition region there are sharp charge reorganizations
causing a dramatic increase in phase sensitivity (delocalization) of up to 4 orders of magnitude.  It is
important to note that the position in parameter space where these reorganizations occur is sample dependent
such that on averaging over many samples only a slight delocalization effect can be observed.  This is found for large disorder $W=7,9$ and for small \textit{repulsive} interactions $U<V$.
Typically two or three distinct charge reorganizations may occur per sample as it is moved between a Mott insulator and an Anderson insulator state.  Attractive interactions favor a more inhomogeneous charge density (forming clusters) and repulsive interactions favor more homogeneous charge density (CDW) as expected.  Of course, this effect will be most pronounced at half filling since for lower densities electrons can avoid each other spatially.

This phenomenon was explained further in terms of avoided level
crossings\cite{schmit2} of the ground state and first-excited state.  However,
to be critical these results are based on small system sizes (20 sites) and
may well be due to finite-size effects.

\subsection{Determining Limits and Accuracy}
As appropriate for any new method, the extent of its validity should be tested before producing results.  There are three points which should be established:
\begin{itemize}
\item The program is working properly.
\item The two methods for gauging the degree of localization yield consistent results.
\item The approximation for eliminating basis states is controlled.
\end{itemize}
The last of these will be the most involved.

\subsubsection{Determining Numerical Limits}
\begin{figure}[ht]
\centering
\includegraphics[width=\mmin{9cm}{\linewidth},clip]{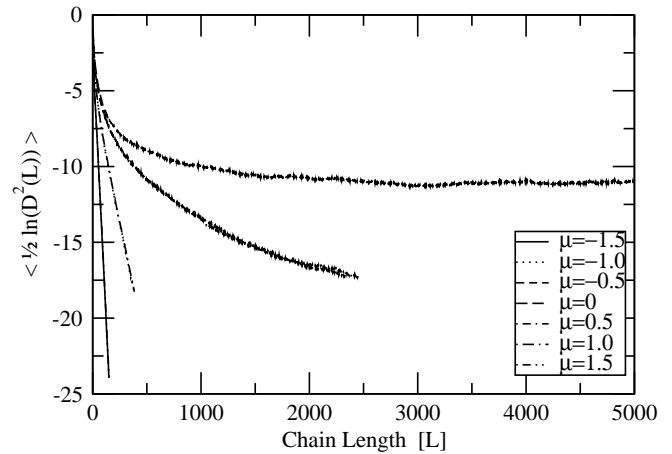}
\caption[Initial results: the length dependence of the phase sensitivity]
{The graph is a plot of the average of $\frac{1}{2}\ln(D^2)$ as a function of chain length $L$ where $D$
 is the phase sensitivity (\ref{eq:ps}).  Averages are
 geometric, that is the mean of $\ln(D^2)$ is found.  The
 chain length is allowed to extend until numerical precision is lost. The $\mu=0$ case was stopped at \mbox{20,000} sites as it had saturated long before.  System parameters are $W=2$, $U=0$ and the energy cutoff is set using $M=10$ (corresponding to a total of approximately 160 basis states per iteration).  The averages were taken over 2000 systems.}
\label{out0}
\end{figure}

The first simulations aimed to determine the broad numerical limits of this new method.  Figure
\ref{out0} shows some typical results where the chain length has been allowed to extend as far as possible.  A
range of values of $\mu$ were used, spread across the band (between $\mu=-2$ and $\mu=2$ in the non-interacting
 case).  The lines pair up, with $\mu=-\mu'$ and $\mu=+\mu'$ giving near identical results and thus demonstrating the
anticipated symmetry in the band.

Straight lines indicate exponential localization, which is seen near the band edge.  However, in the center of the
 band different behavior is observed: the curves show some form of decay which saturates at large chain
 lengths.  In fact, for the $\mu=0$ case, the curve had clearly saturated and so was stopped at \mbox{20 000} sites.

Apart from the $\mu=0$ case, all simulations continued until numerical precision limits were reached. This limit is
encountered when calculating the phase sensitivity.  Off-diagonal elements of the reduced density matrix correspond
to ``scalar product'' type quantities (\ref{eq:rdm}).  Using
this as an analogy, when the two vectors become almost perpendicular the product becomes very small. In this limit, numerical rounding dominates over the physics rendering any results based on this regime meaningless.  Arising out of this, a criterion was devised to automatically halt simulations before this numerically inaccurate region is reached.  Taking
 up the analogy again, when the scalar product divided by the norm of the two vectors is comparable to the
 floating-point precision then only ``noise'' is being calculated.   This
  condition gives
 the upper length limit for linear fits which determine the inverse localization length.  A lower limit was also set in which
  typically the first 20\% of sites were ignored to allow the simulation to ``settle down''.
  
\subsubsection{Particle-Hole Symmetry Test}
The particle-hole symmetry test can be verified by looking at two chains
using the same random distribution of site energies, but with one the negative
of the other (see eqn.~\ref{eq:ehtest}).  On doing so identical results are
obtained.  Thus in the non-interacting case the electron-hole consistency test is convincingly satisfied.

Applying a small electron interaction force $U=0.1$ causes the paired lines to split.  This is also expected. Once the $+\frac{U}{2}\crt{1}\ann{1}+\frac{U}{2}\crt{L}\ann{L}$ correction terms are included, the two lines then collapse on top of each other again.

\subsubsection{Comparison of the two Phase Sensitivity Methods}
\begin{figure}[ht]
\centering
\includegraphics[width=\mmin{8cm}{\linewidth},clip]{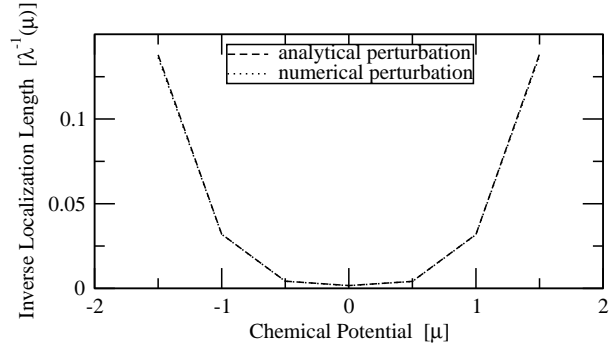}
\caption[Comparison of the two methods for calculating the localization length]
{This graph shows the inverse localization length (obtained from a linear fit) against a range of
values of the chemical potential $\mu$.  The two lines correspond to the two methods of determining localization and
they show very good agreement.  System parameters are $W=2$, $U=0$ with chains allowed to extend to a maximum of 500 sites.
The averages were taken over 2500 systems and the energy cutoff set by $M=10$ (corresponding to a total of approximately
 160 basis states per iteration).  The error bars represent one standard deviation.}
\label{phase2}
\end{figure}

As explained earlier in section \ref{sec:local}, the reduced density matrix method of determining the extent of localization could be verified by calculating the phase sensitivity as a numerical perturbation.  It was found
that the two methods are in good agreement (fig.~\ref{phase2}).  Given the significant computational processing time required for the numerical perturbation method, normally only the reduced density matrix method will be used.

\subsubsection{Reducing the Number of Basis States (Revisited)}
\begin{figure}[ht]
\centering
\includegraphics[width=\mmin{9cm}{\linewidth},clip]{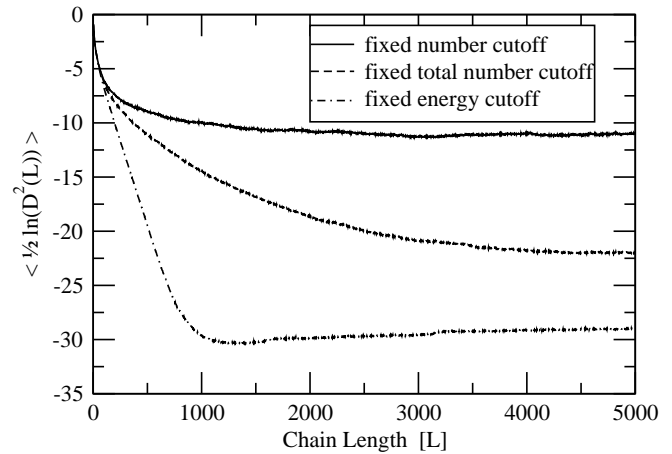}
\caption[Graph demonstrating the effect of changing the cutoff procedure]
{The graph is a plot of the average of $\frac{1}{2}\ln(D^2)$ against chain length $L$.
The parameters are the same as fig.~\ref{out0}, but restricted to the middle of the band ($\mu=0$).
  The only difference between the three curves is the procedure for
discarding basis states.} 
\label{methods}
\end{figure}
The initial results (fig.~\ref{out0}) show that the method fails in the middle
of the band - exponential decay is not observed in the non-interacting case.
This must be due to the Hilbert Space reduction criterion, as it is the only
approximation in the method.  A simple variant on the original procedure was
tried: the \textit{total} number of states for all particle numbers was fixed
rather than using a fixed number for the ground state particle number only.
This simple modification induced a significant change in the results.
Although the decay was still non-exponential, this change clearly yields an improvement toward the expected behavior (fig.~\ref{methods}).

Presumably the key difference between the methods is that using states from
all particle numbers results in an energy cutoff which fluctuates less.
The natural criterion to try next is a fixed energy cutoff.  This can be done by averaging the value of the cutoff using the fixed number of states method.  Then a fit of the cutoff as a function of chain length could be used as a fixed energy cutoff.  It turns out that it is not possible to do this as an absolute cutoff because the ground state energy fluctuates too much.  However it can successfully be done as a fixed energy cutoff relative to the ground state.  When implemented exponential decay is observed in the middle of the band (fig.~\ref{methods}).

We conjecture that this dramatic improvement, resulting from an apparently
innocuous change in cutoff methods, can be explained in terms of energy level statistics. 
Consider the middle of the band with no interactions: electrons should be localized, with states obeying Poisson statistics. One may envisage the system accidentally encountering a higher density of low lying energy states.  According to the original method, the energy cutoff is correspondingly lower.  Thinking in terms of \textit{energy level repulsion}, this would result in a release of ``pressure'' as a larger number of states are removed.  The opposite scenario in which states accidentally spread wider than average may also be considered.  In this case, the cutoff has a smaller effect than normal.  The combined effect is to reduce fluctuations, causing the system to bear more resemblance to a Wigner distribution.  In others words the system tends toward delocalization, consistent with the data on fig.~\ref{methods}.

The second cutoff method implemented worked by fixing the cutoff using states
across all particle numbers.  According to the picture just outlined, the same
effect of dampening fluctuations should still be present, although less severe
because using a greater number of states reduces fluctuations of the cutoff
energy.  This can also be observed in figure~\ref{methods}, where the delocalizing effect is not so strong.

The third procedure for discarding states used a fixed energy cutoff, which is completely uncorrelated to the density of low lying states.  Therefore the delocalizing effect is completely absent.
\begin{figure}[ht]
\centering
\includegraphics[width=\mmin{9cm}{\linewidth},clip]{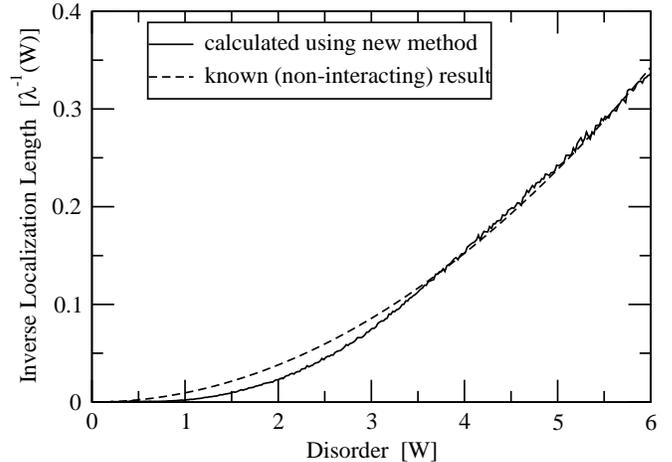}
\caption[Graph showing the dependence of localization upon disorder in the non-interacting limit]
{Graph showing the dependence of inverse localization length upon the disorder when interactions
are turned off.  Results from the new method should correspond to known result for the middle of the band.  Systems were allowed to extend up to 1000 sites, retaining an average of 480 basis states per iteration.  Averages were taken over 1000 disorder realizations.}
\label{avaryw}
\end{figure}
\begin{figure}[ht]
\centering
\includegraphics[width=\mmin{9cm}{\linewidth},clip]{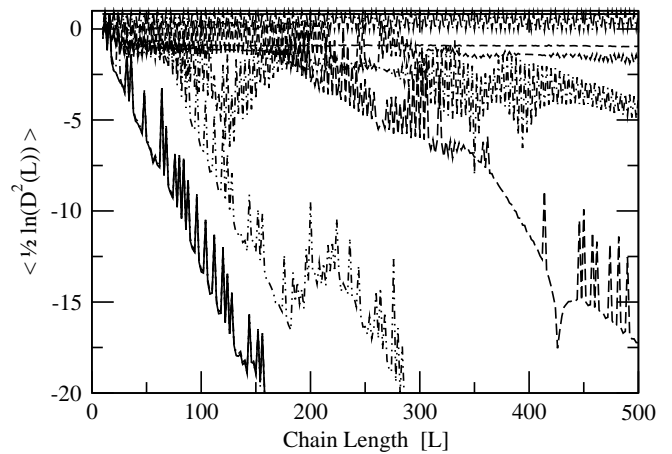}
\caption[Graph showing the failure of the new method for weak disorder]
{Graph showing single sample results for low disorder ($W=0.2$).  The new method fails in this limit.
The only difference between the lines is a slight variation in the energy cutoff.}
\label{loww}
\end{figure}
 
\subsection{Comparison with Non-Interacting Results}
The new fixed energy cutoff may be used to provide further verification by
checking that when interactions are turned off non-interacting results can be
reproduced.  This is particularly important as some well established methods
applied to the two-interacting particle model can fail in this respect (\eg the Transfer Matrix Method \cite{romer2,leadbeater}).

\subsubsection{Dependence on Disorder}
As a test of the accuracy of the method, figure~\ref{avaryw} was produced.  The
calculated results should correspond to the known result (\ref{eq:niresult}).  The new method seems to give a dependence on $W$ greater
than $W^2$.  Values are at least of the correct order of magnitude.  It should be noted, however, that in the clean limit and
for low disorder the new method fails to produce meaningful results.
Figure~\ref{loww} shows a set of results corresponding to the same system but with the energy cutoff slightly varied.  The change in behavior is quite dramatic and dominated by large oscillations.  These effects do not occur for stronger disorder.
  
\subsubsection{Convergence}
\begin{figure}[htp]
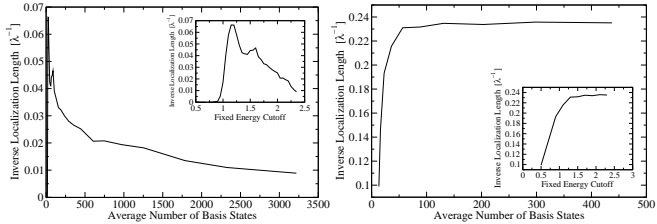

\centering
\includegraphics[width=\mmin{8cm}{0.5\linewidth},clip]{conv1.eps}\hfill
\includegraphics[width=\mmin{8cm}{0.5\linewidth},clip]{conv2.eps}
\caption[Convergence test of the non-interacting case]
{Graph showing the dependence of inverse localization length on the average number of basis
states retained per iteration for the non-interacting case with disorder
$W=2$ (left) and $W=5$ (right).  The insets show this quantity plotted against the actual energy cutoff used.  Such plots can be used to test for convergence.  Data was averaged over 100 systems with
chains extending up to 1000 sites.}
\label{conv1}
\end{figure}
Figure~\ref{conv1} is used to determine whether the method converges to the
known value of the localization length in the middle of the band.  For $W=2$,
eqn.~\ref{eq:niresult} implies $\lambda^{-1} \approx 0.038$.  Hence for
the calculation using the largest matrices the localization length is over
estimated by about a factor \mbox{of 4}.  As anticipated from
fig.~\ref{avaryw}, convergence is much better for $W=5$.  In this case,
eqn.~\ref{eq:niresult} gives $\lambda^{-1} \approx 0.24$.
Figure~\ref{conv1} shows the new method converging at a value close to this result.  One could consider proceeding by just examining interaction effects for $W>4$.  However, interesting physics is expected when disorder and interactions are of similar strengt. Note that the standard deviation, $W/\sqrt{12}$, is a more satisfactory measure of disorder, so that this condition is fulfilled when $W=\sqrt{12}$ and $U=\pm 1$.

\subsubsection{Energy Gaps}
\begin{figure}[ht]
\centering
\includegraphics[width=\mmin{9cm}{\linewidth},clip]{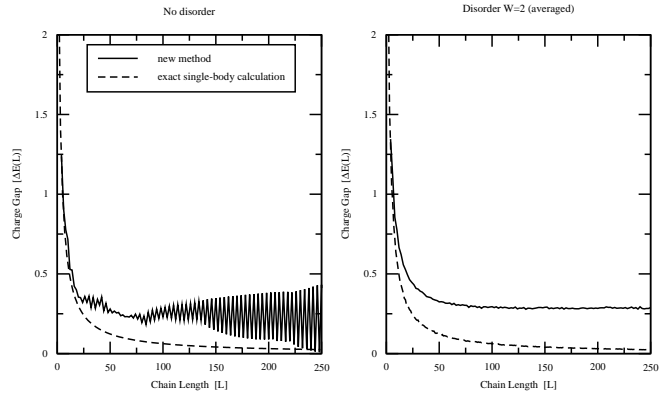}
\caption[Graphs showing the dependence of the charge gap upon system size]
{Graphs showing the dependence of the charge gap $\Delta E$ on chain length.  The black lines are results
from the new method under development and for comparison the 2nd line
represents exact results from a single-body calculation.
The left-hand side shows the clean case and the right-hand side is the average over 1000 systems with disorder $W=2$.  The new fixed energy cutoff was used to remove basis states (with an average of 700 per iteration).}
\label{gapdecay}
\end{figure}
Earlier, in section~\ref{sect:gaps} it was noted that in the non-interacting
limit there is a relation between two of the energy gap definitions
(\ref{eq:gaptest}).  This condition was found to be obeyed provided a large proportion of basis states were retained (\ie only possible for short chain lengths).
 
Also in this limit, the energy gaps should have a specific dependence on length.  This is rigorously known in the clean
limit as $\propto \frac{1}{L}$ for $\Delta E$, $\Delta E_{ph}$ and $\Delta
E_+$.  In fact, when averaged, systems with disorder exhibit the same
behavior.  Figure~\ref{gapdecay} shows how results from the new method
compares with the non-interacting results.  In both cases the decay was only satisfactorily observed for short chain lengths.  The charge gap
decays to a value above zero.  The oscillations arise due to the removal of states which can be established by varying the number of states retained.  This saturation and oscillatory behavior was also observed for the other two gap definitions.

\section{Results}
Despite the unanswered questions, results were successfully obtained using the
\textit{fixed energy} cutoff procedure for eliminating states.  The value for
the cutoff was determined by first using the fixed number procedure for a
small number of systems.  When the average number of basis states is given, it
refers to the number of basis states used with the fixed number procedure in
order to determine the fixed energy cutoff.  In addition note that previous
work by other authors focuses on the case of half-filling, so results presented in this section also examine this particle density.

\subsection{Dependence on Interaction Strength}
\begin{figure}[htb]
\centering
\includegraphics[width=\mmin{7cm}{\linewidth},clip]{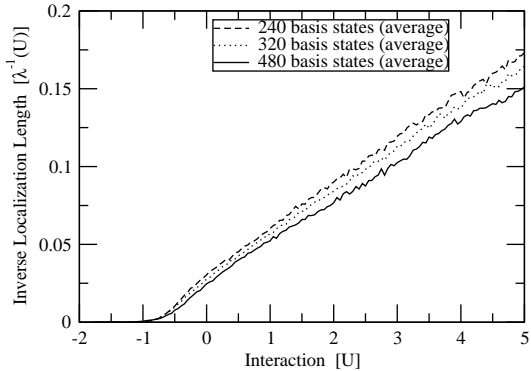}
\caption[Graph showing the dependence of localization upon interaction strength]
{The dependence of localization length on interaction strength.  The three lines correspond to
different energy cutoff values.  Each line is averaged over 1000 systems which are allowed to extend to a maximum
of 1000 lattice sites.  Disorder $W=2$.}
\label{varyu}
\end{figure}
Of central interest is the effect of electron-electron interactions on
localization.  Figure~\ref{varyu} shows the calculated dependency for disorder $W=2$.  Three different energy cutoffs were used corresponding to different
numbers of retained states.

The overall behavior is unambiguous: repulsive interactions enhance the
effect of disorder whereas attractive interactions reduce it.  For $U>0$ the
inverse localization length has an approximately linear relationship to
interaction strength.  Below $U=-1$ the localization length diverges. This
apparently extended regime is anticipated from previous work (section~\ref{sec:ropw}).

Note that due to the ``flattening'' effect (fig.~\ref{psep}) toward the phase separation
the many-body density of states rises rapidly with energy.  For a fixed energy cutoff this means that more states are retained over a greater range of particle numbers.  This reduces computational performance and so the region $-2<U<-1.8$ has not been explored.  Data was obtained down to about $U=-1.8$ and is displayed in fig.~\ref{varyu}.

\begin{figure}[htb]
\centering
\includegraphics[width=\mmin{8cm}{\linewidth},clip]{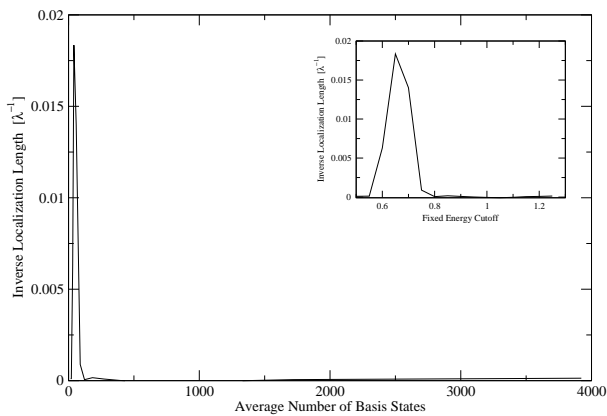}
\caption[Convergence test in the delocalized regime]
{Graph showing the dependence of inverse localization length on the average number of retained basis states in order to test convergence in the delocalized regime.  The value of interaction strength chosen was $U=-1.4$.  The inset shows the dependence on the fixed cutoff energy. The data is averaged over 100 systems with chains allowed to extend to a maximum of 1000 sites.  Disorder $W=2$.}
\label{conv0}
\end{figure}

To explore the convergence, one point in the middle of the delocalized regime
was chosen and its convergence properties were explored.  Figure~\ref{conv0}
demonstrates that, within the computational limits, the extended regime is robust.

\subsection{Disorder-Interaction Phase Space}
\begin{figure}[htb]
\centering
\includegraphics[width=\mmin{12.5cm}{\linewidth},clip]{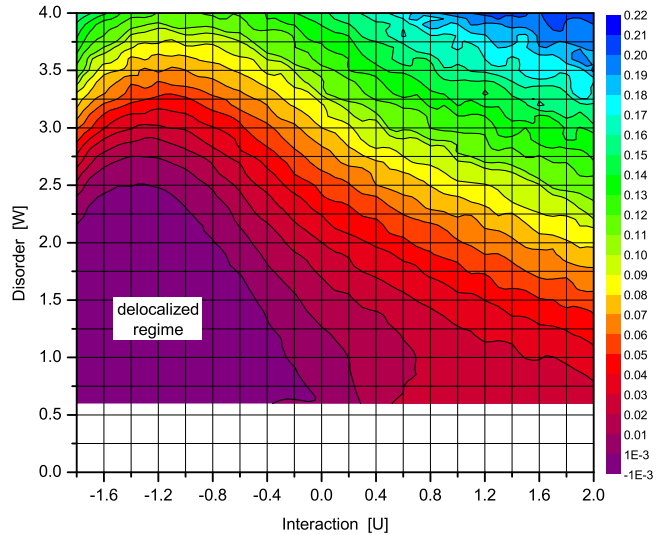}
\caption[Disorder-interaction phase space for the single chain model]
{Disorder-interaction phase space plot for the single chain model at
half-filling.  The spectrum represents the degree of
localization.  The lowest interval corresponds to a localization length greater than 1000 sites.
This contour plot was produced using over 1300 points.  Each point was averaged over 250 systems in which chains were
 allowed to extend to 2000 sites and approximately 240 basis states were retained per iteration.  Data for
 $W<0.6$ is not shown because the method is unreliable for low disorder as discussed earlier.}
\label{varywu}
\end{figure}
Having confirmed the existence of a delocalized regime for attractive interactions, it was natural to attempt to map out the extent of this region.  This was done in disorder-interaction phase space as plotted in fig~\ref{varywu}. The area marked as delocalized corresponds to systems with a localization length greater than 1000 sites.  If the number of systems averaged over were increased then this criterion could be made more stringent.
With this definition it can be seen that the delocalizing effect does not extend beyond $W=2.5$.  The extended region appears to cross the non-interacting line for lower disorder.  However, the method is unable to
produce meaningful results for low disorder.  The phase sensitivity tends to
be dominated by oscillations which appear as the clean limit is approached
(see fig.\ref{loww}).  The (unreliable) results which were obtained indicate that the proximity of the delocalization in fact decreases for lower disorder.  Hence this remains an open question.

\subsection{Determination of Exponents}
\begin{figure}[htb]
\centering
\includegraphics[width=\mmin{9cm}{\linewidth},clip]{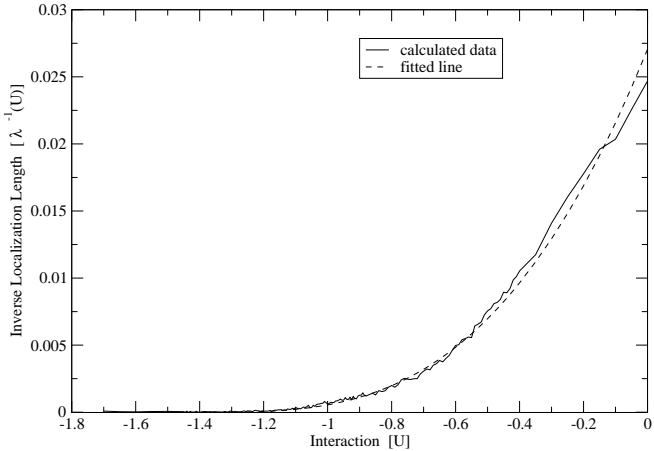}
\caption[Data fitted near the delocalization transition]
{This graph is the same as the black line in fig.~\ref{varyu} but focusing on the transition region.
  The fitted line corresponds to $U_c=-1.375$ and $\nu=3.0$ and shows good agreement.}
\label{exp3}
\end{figure}
It is tempting to describe the change from a localized system to a delocalized system in the language of second-order
 phase transitions.  Although, there is
 no \textit{a priori} reason for believing the transition can be described in this way, it is clear
that a description as a first-order transition is ruled out by the absence of any discontinuities.
Working on this assumption, the obvious quantity to calculate is the exponent defined as
\begin{equation}
\label{eq:fit}
\lambda^{-1} = A (U-U_c)^{\nu} \ \ \ \ U > U_c,
\end{equation}
where $A$ is a coefficient, $\nu$ is the exponent and $U_c$ is the critical interaction strength where the transition occurs.
Two factors make determining this exponent difficult:  Firstly, data can only be used on one side of the transition.
Secondly, the precise value of $U_c$ is unknown and no scaling analysis is available.  Consequently, fitting the data shown on fig.~\ref{varyu} using a log-log plot proved unsatisfactory.  Likewise, attempting to fit all 3 parameters ($A$, $U_c$ and $\nu$) simultaneously was also uncontrolled.  Therefore, an estimate for the critical interaction strength was visually estimated and then the other two parameters could be fitted.  Although, this procedure is less than ideal, it appeared to be the most satisfactory approach.  The estimated range for $U_c$ is $-1.45 \leq U_c \leq -1.3$.   Using a number of values within this range the two other parameters ($A$ and $\nu$) were fitted, giving $\nu \approx 3.0 \pm 0.4$.  This exponent has been plotted with the central value of $U_c$ on fig.~\ref{exp3}.  The fit is not perfect, the deviation may either reflect the absence of a known scaling analysis or that the assumption of a second-order phase transition is incorrect.
 
 However, it is at least clear that away from the transition there is an approximately linear relationship between inverse localization length and the interaction strength. Further, by inspecting fig.~\ref{varyu}, the exponent must therefore be greater than $1$ in the vicinity of the transition.  The determined value for $\nu$ is consistent with that observation.
 
It would be desirable to perform a similar procedure for other values of disorder and to determine the disorder exponent
at a fixed interaction strength (\ie approaching the transition from above on
fig.~\ref{varywu}).  However, the quality of data presently obtained is
inadequate. Schmitteckert~\etal \cite{schmit}, argue that the exponent is
non-universal.

\subsection{Summary}
We have presented a method of studying disordered and interacting
quasi-1-dimensional systems which combines aspects of the transfer matrix and
DMRG approaches.  While the method works well and is able to study
significantly larger systems than have been achieved hitherto, there is still
room for improvement.  In particular the strategy for reducing the Hilbert
space and compensating for the side effects of the reduction is still too
simplistic.  It would be useful to understand why the method fails so
dramatically for low disorder. Nevertheless the method is generalizable to
more complex problems such as the Hubbard model or strips of finite width.
It could eventually be possible to combine such an approach with finite size
scaling in order to study the metal-insulator transition.

As a first application of our method we have presented results on spinless
fermions in 1D.  There is qualitative agreement with previous work: repulsive
interactions increase the effect of disorder and attractive interactions have
the opposite effect.  We have mapped out the delocalized regime and
found some disagreement with previous work. According to our results, DMRG
studies \textit{under} estimate this region by a factor of 2 and an earlier
study \textit{over} estimates it by a factor of 2.  We have also made a first
estimate of the critical exponent of this transition, but our data is not yet
sufficient to test its universality.

\bibliography{local}

\end{document}